\documentclass[aps,prl,twocolumn,superscriptaddress,floatfix,longbibliography]{revtex4-2}

\usepackage{graphicx}
\usepackage{braket}
\usepackage{bm}
\usepackage{upgreek}
\usepackage{tabularx}
\usepackage[normalem]{ulem}
\usepackage[utf8]{inputenc}
\usepackage{amsmath}
\usepackage{amsfonts}
\usepackage{amssymb}
\usepackage{todonotes}
\usepackage{comment}
\usepackage{siunitx}
\usepackage{mathtools}
\usepackage{array}
\usepackage{physics}
\usepackage{booktabs}

\usepackage{hyperref}

\begin{document}
\title{Signatures of exciton orbits in quantum mechanical recurrence spectra of Cu$_2$O}
\author{Jan Ertl}
\affiliation{Institut für Theoretische Physik I, Universität
  Stuttgart, 70550 Stuttgart, Germany}
\affiliation{Experimentelle Physik 2, Technische Universität Dortmund,
  44221 Dortmund, Germany}
\author{Michael Marquardt}
\author{Moritz Schumacher}
\author{Patric Rommel}
\author{Jörg Main}
\email[Email: ]{main@itp1.uni-stuttgart.de}
\affiliation{Institut für Theoretische Physik I, Universität
  Stuttgart, 70550 Stuttgart, Germany}
\author{Manfred Bayer}
\affiliation{Experimentelle Physik 2, Technische Universität Dortmund,
  44221 Dortmund, Germany}

\date{\today}

\begin{abstract}
The seminal work by T.~Kazimierczuk et al.\ [Nature \textbf{514}, 343 (2014)]
has shown the existence of highly excited exciton states in a regime,
where the correspondence principle is applicable and quantum mechanics
turns into classical mechanics, however, any interpretation of exciton
spectra based on a classical approach to excitons is still missing.
Here, we close this gap by computing and comparing quantum mechanical
and semiclassical recurrence spectra of cuprous oxide.
We show that the quantum mechanical recurrence spectra exhibit peaks,
which, by application of semiclassical theories and a scaling
transformation, can be directly related to classical periodic exciton orbits.
The application of semiclassical theories to exciton physics requires
the detailed analysis of the classical exciton dynamics, including
\emph{three-dimensional} orbits, which strongly deviate from
hydrogenlike Keplerian orbits.
Our findings illuminate important aspects of excitons in
semiconductors by directly relating the quantum mechanical
band-structure splittings of excitons to the corresponding classical
exciton dynamics.
\end{abstract}

\maketitle

Excitons are atom-like states in semiconductors formed by an electron
and a positively charged hole.
They are created by exciting an electron from the valence band into
the conduction band, where the electron forms a bound hydrogenlike
state with the hole remaining in the valence band~\cite{Gross1956,Luttinger56}.
Since the experimental observation of giant Rydberg excitons with
principal quantum numbers up to $n=25$ in Cu$_2$O by Kazimierczuk
et al.~\cite{GiantRydbergExcitons} the exciton physics of cuprous oxide
has attracted a strongly increasing interest, both experimentally
\cite{GiantRydbergExcitons,Schoene2016,SchoeneLuttinger,ObservationHighAngularMomentumExcitons}
and theoretically 
\cite{Schoene2016,SchoeneLuttinger,ObservationHighAngularMomentumExcitons,ImpactValence,frankevenexcitonseries}.
In particular, the impact of the valence band structure causes
significant deviations of the exciton spectra from
a simple hydrogenlike model, which have now been investigated in great
detail~\cite{Schoene2016,SchoeneLuttinger,ObservationHighAngularMomentumExcitons,ImpactValence}.

For the hydrogen atom the connection of Rydberg spectra to classical
Keplerian orbits is well established by the Bohr-Sommerfeld model.
Semiclassical trace formulas \cite{Berry76,Gut90}
provide the link between quantum spectra and classical dynamics of
both regular and chaotic systems, and are the foundation for the
understanding of level-spacing dynamics \cite{Haake2018}.
For example, the diamagnetic Kepler problem has served as a prototype
system for the study of \emph{quantum chaos}, i.e., the effects of a
classical chaotic dynamics on quantum spectra~\cite{Fri89,Has89}.
Due to the additional spin degrees of freedom in the
semiconductor and the strong, non-negligible spin-orbit interaction
the existence of an exciton dynamics for quasiparticles similar as for
electrons in hydrogen atoms is not obvious.
A first step on a semiclassical description of Rydberg excitons in
Cu$_2$O has been made in Ref.~\cite{Ertl2020} by proposing a classical
model using an adiabatic approach that separates the fast spin
dynamics and the slow electron-hole dynamics.
By constructing action variables for the exciton dynamics in certain
symmetry planes of the crystal allowed energy regions for the
existence of exciton states could be obtained.
This gives a first hint for the validity of a semiclassical approach,
however, a direct verification of a classical exciton dynamics is
still missing.
By computing the quantum mechanical recurrence spectra and comparing
them with the classical and semiclassical results, we reveal,
the existence and meaningfulness of a classical exciton dynamics.
Direct signatures of exciton orbits obtained in quantum mechanical
exciton recurrence spectra provide an intuitive picture for the
understanding of excitons in Cu$_2$O.

The investigation of the phase space topology with similar methods as
introduced by Gekle et al.~\cite{Gekle2006a,Gekle2007a} for the
hydrogen atom in crossed electric and magnetic fields, reveals a
mostly regular or near-integrable exciton dynamics with periodic
orbits on one- to three-dimensional tori.
Here, we establish the connection between the
fine-structure splitting of excitons in quantum spectra of the yellow
series of Cu$_2$O and the corresponding classical exciton dynamics.
According to semiclassical theories~\cite{Berry76,Gut90},
the frequencies of the oscillations are determined by the action 
of the periodic orbits, and the amplitudes are related to
stability properties of the orbits at a given energy.
Performing a Fourier transform of the density of states from energy to 
time domain results in a recurrence spectrum which exhibits peaks at periods that 
can be assigned to classical orbits~\cite{a_SMErtl2022}. 
For systems exhibiting an appropriate scaling property the peaks
in the recurrence spectrum become sharp $\delta$ peaks~\cite{main1999a}.
To this aim we apply a scaling technique and calculate the classical 
dynamics of the system.
The obtained periodic exciton orbits perfectly explain the observed
structures in the quantum mechanical recurrence spectra, and
thus provide a deeper physical understanding of excitons in semiconductors.

A full description of excitons in cuprous oxide needs to consider the
cubic $O_{\mathrm{h}}$ symmetry of the system.
Introducing relative and center-of-mass coordinates for the electron
and hole and neglecting the center-of-mass momentum, the Hamiltonian
for excitons in cuprous oxide is given by
\cite{Lipari1977,Uihlein1981,ImpactValence}
\begin{equation}
  H = E_{\mathrm{g}} + H_{\mathrm{kin}}(\boldsymbol{p},\boldsymbol{\hat{I}},\boldsymbol{\hat{S}}_{\mathrm{h}})
  -\frac{e^2}{4\pi\varepsilon_0\varepsilon|\boldsymbol{r}|} + H_{\mathrm{SO}} \, ,
\label{eq:H}
\end{equation}
with the relative coordinates $\boldsymbol{r}$ and momenta $\boldsymbol{p}$
and the vector operators $\boldsymbol{\hat{I}}$, $\boldsymbol{\hat{S}}_{\mathrm{h}}$
for angular momenta $I=1$ and $S_\mathrm{h}=1/2$.
Here, the first term $E_{\mathrm{g}}=2.17208\,$eV is the gap energy
between the uppermost valence band and the lowest conduction band
\cite{GiantRydbergExcitons}.
The second term
\allowdisplaybreaks
\begin{align}
  &H_{\mathrm{kin}} (\boldsymbol{p},\boldsymbol{\hat{I}},\boldsymbol{\hat{S}}_{\mathrm{h}})
    = \frac{\gamma'_1}{2m_0} \boldsymbol{p}^2
    +\frac{1}{2\hbar^2m_0} \big[4\gamma_2\hbar^2\boldsymbol{p}^2\nonumber\\[1ex]
  &-6\gamma_2(p^2_1\hat{I}^2_1+{\rm c.p.})
    -12\gamma_3(\{p_1,p_2\}\{\hat{I}_1,\hat{I}_2\}+{\rm c.p.})\nonumber\\[1ex]
  &-12\eta_2(p^2_1\hat{I}_1\hat{S}_{\mathrm{h1}}+{\rm c.p.})
    +2(\eta_1+2\eta_2)\boldsymbol{p}^2(\boldsymbol{\hat{I}}\cdot\boldsymbol{\hat{S}}_{\mathrm{h}})\nonumber\\[1ex]
  &-12\eta_3(\{p_1,p_2\}(\hat{I}_1\hat{S}_{\mathrm{h2}}
    +\hat{I}_2\hat{S}_{\mathrm{h1}})+{\rm c.p.})\big] \, ,
\label{eq:H_kin}
\end{align}
accounts for the kinetic energy of the electron and hole.
It includes the cubic band structure, described by the quasispin
$\boldsymbol{\hat{I}}$ and the hole spin $\boldsymbol{\hat{S}}_{{\mathrm h}}$
with their components $\hat{I}_i$, $\hat{S}_{{\mathrm h}i}$ as well as
the components of the momentum $p_i$.
Furthermore, $m_0$ is the free-electron mass, $\{a,b\}=\frac{1}{2}(ab+ba)$
denotes the symmetrized product, c.p.\ stands for cyclic permutation,
$\gamma_i$ and $\eta_i$ are the Luttinger parameters~\cite{SchoeneLuttinger},
and $\gamma'_1= \gamma_1 + m_0/m_{\mathrm{e}}=2.77$.
The third term in Eq.~\eqref{eq:H} is the screened Coulomb potential
with the dielectric constant $\varepsilon=7.5$.
The fourth term in Eq.~\eqref{eq:H} is the spin-orbit term
\begin{equation}
  H_{\mathrm{SO}}=\frac{2}{3}\Delta
  \left(1+\frac{1}{\hbar^2}\boldsymbol{\hat{I}}\cdot\boldsymbol{\hat{S}}_{\mathrm{h}}\right)\, ,
\label{eq:H_SO}
\end{equation}
where $\Delta=0.131\,$eV is the spin-orbit coupling \cite{SchoeneLuttinger}.
In our computations we use the same material parameters as
given in Ref.~\cite{Rommel2020Green}, but neglect central-cell
corrections~\cite{frankevenexcitonseries,Farenbruch2020a,Rommel2021a},
which can be justified for high principal quantum numbers in the
semiclassical limit~\cite{b_SMErtl2022}.
For a given energy the classical dynamics of the yellow exciton series
can be calculated by using the adiabatic approach introduced
in~\cite{Ertl2020,b_SMErtl2022}.

Without the spin-orbit term~\eqref{eq:H_SO} the Hamiltonian~\eqref{eq:H}
does not depend on the energy when multiplied by
$n_\mathrm{eff}^2$ and performing a scaling transformation
$\boldsymbol{r} = n_{\mathrm{eff}}^2\tilde{\boldsymbol{r}}$,
$\boldsymbol{p} = n_{\mathrm{eff}}^{-1}\tilde{\boldsymbol{p}}$,
with the effective quantum number $n_{\mathrm{eff}}$ given by
$n_{\mathrm{eff}}^2\equiv E_{\mathrm{Ryd}}/(E_{\mathrm{g}}-E)$
with $E_{\mathrm{Ryd}}=13.6$eV$/(\gamma_1' \varepsilon^2)\approx
87$meV the Rydberg energy of cuprous oxide,
which means that the classical dynamics is the same for all values of
$n_\mathrm{eff}$.
The non-scaled action $S$ is  
connected to the scaled value $\tilde S$ by a simple linear scaling
$S_\mathrm{po}(n_\mathrm{eff})=\tilde{S}_\mathrm{po} n_\mathrm{eff}$.
In semiclassical theories the 
density of states for systems with such a scaling property can be
expressed as a Fourier series in the scaled action
$\tilde S_{\mathrm{po}}$~\cite{main1999a,a_SMErtl2022},
\begin{equation}
  \varrho(n_{\mathrm{eff}}) = \varrho_0(n_{\mathrm{eff}})
  +\Re \sum_{\mathrm{po}} {\cal A}_{\mathrm{po}}
    \exp(i\tilde S_{\mathrm{po}}n_{\mathrm{eff}}/\hbar) \, ,
\label{eq:rho_sc}
\end{equation}
with $\varrho_0(n_{\mathrm{eff}})$ the average density of states.
The sinusoidal fluctuations of the density are related to the periodic
orbits (po) of the classical system with ${\cal A}_{\mathrm{po}}$ and
$\tilde{S}_\mathrm{po}$ the amplitude (including the Maslov index)
and the scaled action of the orbits, respectively.

To recover the scaling property for the Hamiltonian we introduce an 
energy-dependent coupling parameter $\tilde\Delta$, i.e.,
\begin{equation}
  \Delta \to \tilde \Delta = \frac{n_0^2}{n_{\mathrm{eff}}^2} \Delta \, ,
\label{eq:Delta_scal}
\end{equation}
with a fixed parameter $n_0$.
Note that the replacement~\eqref{eq:Delta_scal} is not possible in an
experiment, however, a tunable spin-orbit coupling $\Delta$ has
already been used for the theoretical investigation of the exchange
interaction in the yellow exciton series~\cite{Rommel2021a}.
The classical dynamics is then that of $n_\mathrm{eff}=n_0$.
Using the adiabatic approach~\cite{Ertl2020,b_SMErtl2022} and choosing
the lowest-lying energy surface in momentum space
corresponding to the yellow exciton series classical exciton orbits
can be obtained by numerical integration of Hamilton's equations of
motion for the relative coordinates and momenta.
In most parts of the phase space we observe a regular dynamics of the
excitons on one- to three-dimensional tori.
Periodic orbits on these tori can be described by one to three
integer winding numbers $M_i$.
The number of winding numbers for the three-dimensional 
orbits can be reduced to an effective two-dimensional description by 
two winding numbers $M_1$ and $M_2$~\cite{b_SMErtl2022}.
The corresponding action variables are $J_1$ and $J_2$.
At given energy $E$ they are related by the function
$g_E(J_1)=J_2$, which can be used to compute the semiclassical
amplitudes of periodic orbits on resonant tori~\cite{Berry76,Tomsovic95}.
In addition the stability eigenvalues $\lambda_{\mathrm{po}}$, which
describe the linearized response of a periodic orbit to a small
perturbation are required for the calculation of the semiclassical
amplitudes in Eq.~\eqref{eq:rho_sc}.
Since only one pair of stability eigenvalues shows deviations from the
integrable behavior for the majority of orbits we use a mixed approach
combining the amplitudes of the Berry-Tabor formula for a
two-dimensional system~\cite{Berry76,Tomsovic95} with the contribution of the
stability eigenvalues $\lambda_{\mathrm{po}}$ and $1/\lambda_{\mathrm{po}}$
for the unstable direction from Gutzwiller's trace
formula~\cite{Gut90}, resulting in the equation
\begin{equation}
|{\cal A}_{\mathrm{po}}|=
\frac{1}{\pi\hbar}
      \frac{1}{\sqrt{\abs{\lambda_{\mathrm{po}}+1/ \lambda_{\mathrm{po}}-2}}}
      \frac{\tilde{S}_{\mathrm{po}}}{\sqrt{\hbar M_2^3 \abs{g''_E}}}
\label{eq:amp_sc}
\end{equation}
for the periodic-orbit amplitudes~\cite{b_SMErtl2022}.
For the computation of the amplitudes of the isolated nearly circular
orbits we resort directly to Gutzwiller's trace formula.

In the quantum mechanical case the operators for position and momentum read
$\hat{\tilde{\boldsymbol{r}}} = \tilde{\boldsymbol{r}}$,
$\hat{\tilde{\boldsymbol{p}}} = -i\hbar_{\mathrm{eff}}\nabla_{{\tilde r}}$,
with $\hbar_{\mathrm{eff}}=\hbar/n_{\mathrm{eff}}$ an effective Planck
constant~\cite{c_SMErtl2022}.
The Schrödinger equation for cuprous oxide is now transformed to the
generalized eigenvalue problem
\begin{equation}
  \left[\frac{e^2}{4\pi\varepsilon_0\varepsilon|\tilde{\boldsymbol{r}}|}
    - n_0^{2}H_{\mathrm{SO}} - E_{\mathrm{Ryd}} \right] |\Psi \rangle
  = \hbar_{\mathrm{eff}}^2 H_{\mathrm{kin}} |\Psi \rangle
\label{eq:H_scal}
\end{equation}
for the squared effective Planck constant, i.e.\ $\lambda=\hbar_{\mathrm{eff}}^2$,
and thus the effective quantum number $n_{\mathrm{eff}}$.
Eq.~\eqref{eq:H_scal} is solved numerically by using a complete set of
basis states $|N L J F M_F \rangle$ with Coulomb-Sturmian radial
functions $U_{NL}(r)$~\cite{c_SMErtl2022}.

The decisive point of the scaling is that the eigenvalues in
Eq.~\eqref{eq:H_scal} correspond to the effective Planck constant
$\hbar_{\mathrm{eff}}=\hbar/n_{\mathrm{eff}}$, i.e., the eigenstates
approach the semiclassical limit with increasing eigenvalues
$n_{\mathrm{eff}}$, but the classical exciton dynamics corresponding to the
spectrum does not depend on this effective Planck constant, and thus
stays the same for all states of the scaled spectrum.
The classical exciton dynamics is that of the non-scaled
Hamiltonian~\eqref{eq:H} ($n_{\mathrm{eff}}=n_0$ in
Eq.~\eqref{eq:Delta_scal}) at energy $E=E_{\mathrm{g}}-E_\mathrm{Ryd}/n_0^2$, and
is thus controlled via the parameter $n_0$ in
Eqs.~\eqref{eq:Delta_scal} and \eqref{eq:H_scal}.
The fluctuations of the scaled quantum spectra obtained from
Eq.~\eqref{eq:H_scal} can be analyzed by Fourier transform in the
variable $n_{\mathrm{eff}}$ and, via the semiclassical result~\eqref{eq:rho_sc},
should provide $\delta$ peaks at frequencies given by the scaled actions
$\tilde S_{\mathrm{po}}$ of the periodic orbits of the corresponding
classical exciton dynamics.

\begin{figure}
  \includegraphics[width=\columnwidth]{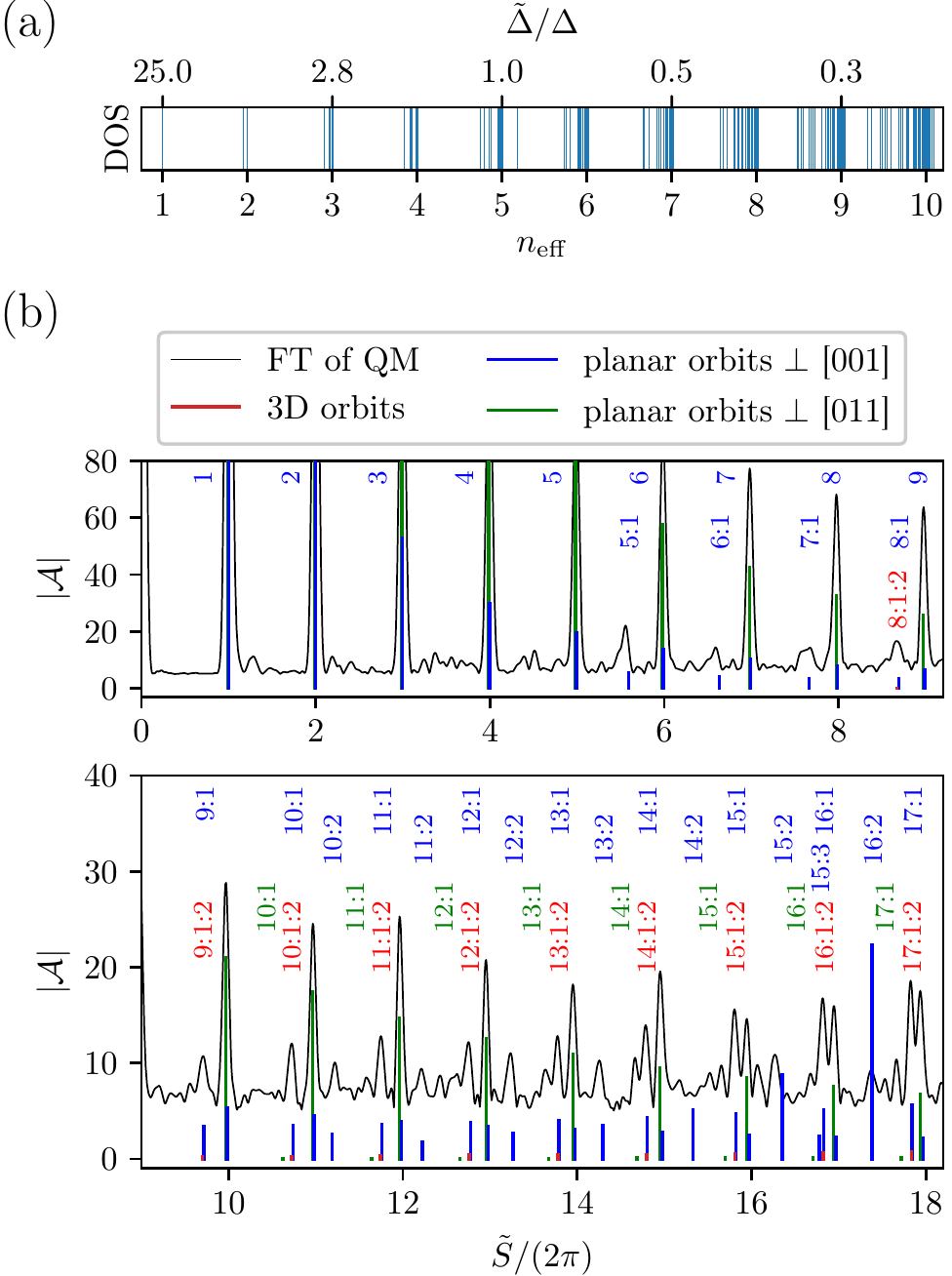}
  \caption{(a) Part of the scaled quantum mechanical density of
    states for $n_0=5$.  (b) Quantum mechanical exciton recurrence
    spectrum (black solid line, with zero line shifted for better
    visibility) obtained by Fourier transform (FT) of the density of
    states shown in (a) and the semiclassical recurrence spectrum
    (colored bars).  The peaks corresponding to one or multiple
    repetitions of nearly circular orbits are labeled with a single
    winding number $M_1$.  Two winding numbers $M_1{:}M_2$ indicate
    planar orbits in one of the two different symmetry planes of the
    crystal, and fully three-dimensional orbits are marked by three
    winding numbers $M_1{:}M_2{:}M_3$.
    The observed structures agree very well with
    the semiclassical results.}
  \label{fig:semiclassical}
\end{figure}

In the following all parameters are given in exciton-Hartree units
which are obtained by setting $\hbar = e = m_{\mathrm{0}} / \gamma_1'
= 1 / (4\pi\varepsilon_{\mathrm{0}} \varepsilon) = 1$.
For the presentation of the results we focus on $n_0=5$, i.e.,
a principal quantum number, which is high enough that the adiabatic
approach is valid, but low enough that the secular motion of the
classical exciton orbits, which decreases with increasing $n_0$, is
sufficiently fast~\cite{b_SMErtl2022}.
Numerical diagonalization of the generalized eigenvalue problem
\eqref{eq:H_scal} provides the scaled quantum mechanical spectrum for
$n_{\mathrm{eff}}$ displayed in Fig.~\ref{fig:semiclassical}(a)
showing nicely the fine structure splittings.
Due to the scaling property introduced in Eq.~\eqref{eq:Delta_scal} the
spectrum differs from the physical (non-scaled) spectrum.
The ratio of the scaled spin-orbit splitting $\tilde \Delta$ to the
physical value $\Delta$ is shown in the upper axis of
Fig.~\ref{fig:semiclassical}(a).
In the vicinity of $n_{\mathrm{eff}} \to n_0$ the scaled spin-orbit
splitting and the physical value coincide leading to a good agreement
of physical and scaled spectra in this energy range.
Contrary to the physical spectrum, the scaled spectrum can be
understood directly in terms of classical orbits, because the
semiclassical density of states for the scaled systems is given
as a Fourier series with the semiclassical amplitudes
${\cal A}_{\mathrm{po}}$ at positions $\tilde S_{\mathrm{po}}$. 

The scaled quantum spectrum shown in Fig.~\ref{fig:semiclassical}(a)
is a sum of $\delta$ distributions, and thus the Fourier transform (FT)
can easily be carried out analytically.
The resulting Fourier spectrum is presented as black solid line in
Fig.~\ref{fig:semiclassical}(b).
It exhibits a large number of sharp peaks with increasing density as a
function of $\tilde S$.
The peaks should approach $\delta$ functions, i.e., become infinitely
narrow for the Fourier transform of an infinitely long scaled quantum
spectrum with $n_{\mathrm{eff}} \to \infty$.
However, for finite length, resulting from the diagonalization of the
truncated generalized eigenvalue problem \eqref{eq:H_scal}, the peaks
are broadened and side peaks may occur.
To suppress these features we use a Gaussian window function as an
envelope when performing the Fourier transform~\cite{c_SMErtl2022}.
\begin{figure}
  \includegraphics[width=\columnwidth]{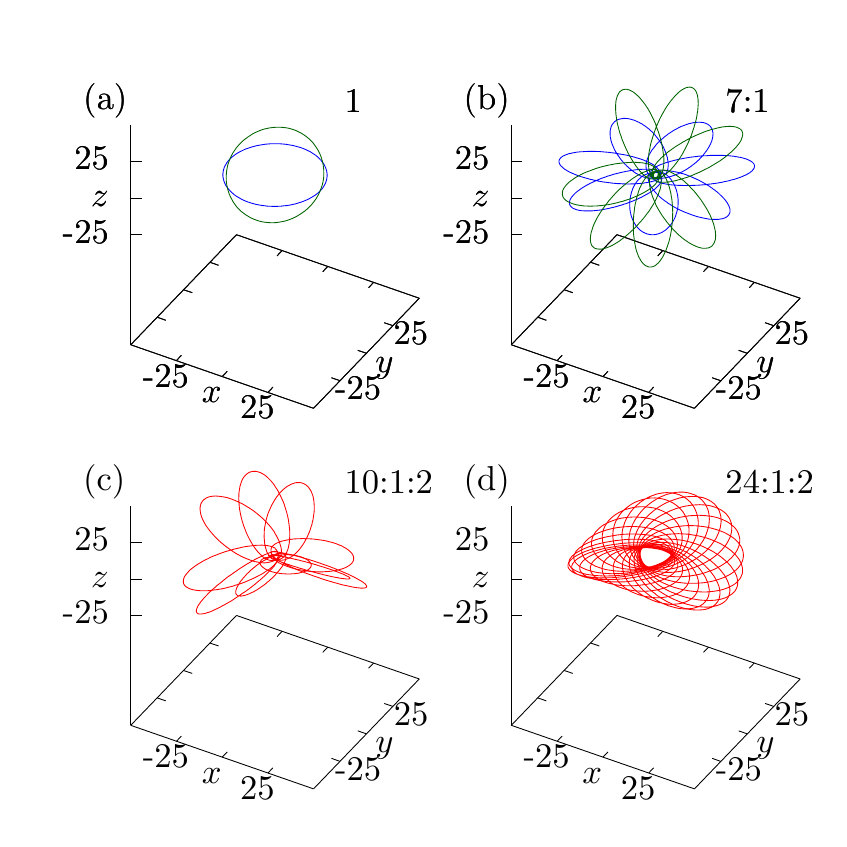}
  \caption{(a) Nearly circular orbits with winding number $M_1=1$ and
    (b) planar orbits with winding numbers $M_1{:}M_2=7{:}1$ in the
    two different symmetry planes of the crystal.  (c) and (d): Two
    examples of fully three-dimensional orbits with winding numbers
    $M_1{:}M_2{:}M_3$.
    The colors are the same as in Fig.~\ref{fig:semiclassical}.}
  \label{fig:Orbits}
\end{figure}

For comparison, the semiclassical results are shown in
Fig.~\ref{fig:semiclassical}(b) as colored peaks at the positions
$\tilde S_{\mathrm{po}}/2\pi$ of the periodic orbits.
The peak heights mark the absolute values $|{\cal A}_{\mathrm{po}}|$
of the semiclassical amplitudes.
The peaks are labeled by the one to three winding numbers $M_i$ of the
corresponding periodic orbits moving on one- to three-dimensional
tori.
As can be seen in Fig.~\ref{fig:semiclassical}(b), the quantum
mechanical and semiclassical exciton recurrence spectra agree very well.
At low action $(\tilde S/2\pi\lesssim 5)$ the peaks solely belong to
one or multiple recurrences of the two shortest periodic exciton
orbits, viz.\ the nearly circular orbits in the planes perpendicular
to the $[001]$ and $[011]$ axes, moving on one-dimensional tori
labeled by a single winding number $M_1=1,2,3,\dots$.
These orbits are shown in Fig.~\ref{fig:Orbits}(a).
At higher action $(\tilde S/2\pi\gtrsim 5)$ the recurrence spectrum
becomes more and more complicated due to the appearance of additional
peaks belonging to exciton orbits on two-dimensional tori located in
the planes perpendicular to the $[001]$ and $[011]$ axes (marked by
two winding numbers $M_1{:}M_2$) or fully three-dimensional orbits
with winding numbers $M_1{:}M_2{:}M_3$.
The 2D orbits with $M_1{:}M_2=7{:}1$ in the symmetry planes of the
crystal are shown in Fig.~\ref{fig:Orbits}(b).
Two fully 3D exciton orbits are illustrated in
Fig.~\ref{fig:Orbits}(c) and (d).

\begin{figure}
  \includegraphics[width=\columnwidth]{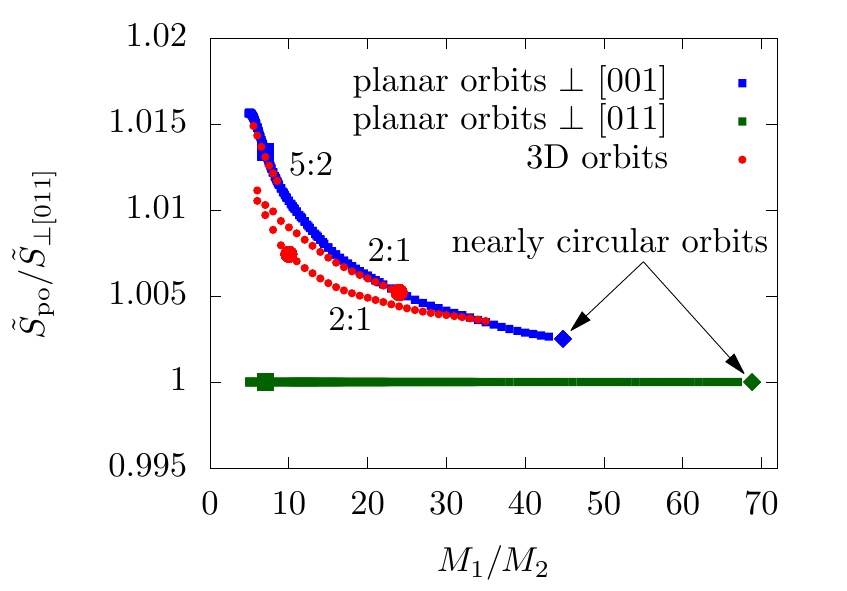}
  \caption{Actions $\tilde S_{\mathrm{po}}$ of periodic orbits as
    function of the ratio of winding numbers $M_1/M_2$.  The actions
    are normalized by the actions $\tilde S_{\perp [011]}$ of the
    corresponding orbits with the same winding numbers $M_1$ and $M_2$
    in the plane perpendicular to the $[011]$ axis.  The two-dimensional
    orbits approach the action of the nearly circular orbit (indicated
    by rhombi) of the corresponding plane with increasing $M_1/M_2$.
    Some three-dimensional orbits with marked ratio $M_3{:}M_2$  are
    located in the area enclosed by orbits in the two different
    symmetry planes of the $O_{\mathrm{h}}$ group. The orbits shown in
    Fig.~\ref{fig:Orbits} are highlighted by larger symbols.}
  \label{fig:Actions}
\end{figure}
The structure of the classical exciton dynamics is illustrated in more
detail in Fig.~\ref{fig:Actions}, where the classical action
$\tilde S_{\mathrm{po}}$ of the period orbits is shown as function of
the ratio of the winding numbers $M_1/M_2$.
For better visibility, the actions are normalized by the actions
$\tilde S_{\perp [011]}$ of the corresponding orbits with the same
winding numbers $M_1$ and $M_2$ in the plane perpendicular to the
$[011]$ axis.
Therefore, by construction, the periodic orbits in the plane
perpendicular to the $[011]$ axis are located on the straight line at
$\tilde S_{\mathrm{po}}/\tilde S_{\perp [011]}=1$.
These orbits lie on a 2D torus in phase space.
With increasing ratio $M_1/M_2$ they converge at $M_1/M_2 \approx 68.8$ to
the nearly circular orbit in the symmetry plane perpendicular to $[011]$.
This point in Fig.~\ref{fig:Actions} thus represents a limiting 1D
torus in phase space indicated by the green rhombus.
In a similar way the periodic orbits perpendicular to the $[001]$ axis,
lying on a different 2D torus in phase space, are located on the upper
line in Fig.~\ref{fig:Actions} with the limiting nearly circular orbit on
a 1D torus at $M_1/M_2 \approx 44.8$ shown as blue rhombus.
In between the orbits on the two limiting 2D tori the periodic orbits
on the 3D tori are located.
Subsets of these orbits with winding number $M_2=1$ or $2$ and ratios
$M_3{:}M_2=2{:}1$ or $5{:}2$ are marked by red dots in
Fig.~\ref{fig:Actions}.
The three-dimensional orbits fill the area between the limiting 2D
tori more densely when longer periodic orbits with more complicated
ratios of the winding numbers are considered.
As can be seen in Fig.~\ref{fig:Actions}, the classical action of
periodic orbits with the same winding numbers $M_1$ and $M_2$ differ
by less than $2$\%.
This causes a clustering of several orbits in the recurrence spectra
in Fig.~\ref{fig:semiclassical}(b).
In these clusters the peaks at the highest action belong to the
two-dimensional orbits in the (mostly) stable symmetry plane
perpendicular to the $[001]$ axis and its equivalents.
Typically, these peaks exhibit the highest semiclassical amplitude
within the cluster.
At slightly lower action one finds unstable three-dimensional orbits,
and the peak with lowest action of the cluster belongs to a
two-dimensional orbit in the plane perpendicular to the $[011]$ axis.

In summary, we have found signatures of exciton orbits in
quantum mechanical recurrence spectra of cuprous oxide.
We have revealed the classical phase space structure of
yellow excitons in cuprous oxide and observed recurrence peaks in
Fourier transform quantum spectra, which, by application of
semiclassical theories can be directly related to two-dimensional
periodic orbits in symmetry planes of the crystal or fully
three-dimensional periodic orbits.
The results have been obtained by using an adiabatic approach for the
classical exciton dynamics and by application of a scaling technique
to the quantum spectra.
Considering these approximations and that the Bohr-Sommerfeld 
model already fails to predict the energy levels of the helium atom,
it is remarkable that a classical picture is capable of describing the
spectral features of excitons in Cu$_2$O.
Here, we have focused
on the dynamics of Rydberg excitons with principal
quantum number $n=5$ in the (non-scaled) Cu$_2$O crystal.
The analysis will be extended to other dynamics
regimes by varying $n_0$.
It will also be interesting to investigate the classical and
semiclassical dynamics of magnetoexcitons~%
\cite{frankmagnetoexcitonscuprousoxide,frankmagnetoexcitonsbreak}
in cuprous oxide.
Furthermore, the classical model intrinsically exhibits a dipole
moment which could provide a starting point to describe and better
understand interactions between Rydberg excitons such as scattering
processes between Rydberg excitons~%
\cite{hefetz1985observation,shumway2001quantum}, the Rydberg blockade~%
\cite{heckotter2021asymmetric}, and the possible existence of an
exciton molecule \cite{bassani1976biexciton} in analogy to Rydberg
molecules \cite{bendkowsky2009observation,junginger2012quantum}.

\acknowledgments
This work was supported by Deutsche Forschungsgemeinschaft (DFG)
through Grant No.~MA1639/16-1 and through the International
Collaborative Research Centre (ICRC) TRR 160 (project A8).

\bibliography{paper}

\end{document}


\title{Supplemental material for ``Signatures of exciton orbits in quantum mechanical
  recurrence spectra of Cu$_2$O''}
\author{Jan Ertl}
\affiliation{Institut für Theoretische Physik I, Universität
  Stuttgart, 70550 Stuttgart, Germany}
\affiliation{Experimentelle Physik 2, Technische Universität Dortmund,
  44221 Dortmund, Germany}
\author{Michael Marquardt}
\author{Moritz Schumacher}
\author{Patric Rommel}
\author{Jörg Main}
\email[Email: ]{main@itp1.uni-stuttgart.de}
\affiliation{Institut für Theoretische Physik I, Universität
  Stuttgart, 70550 Stuttgart, Germany}
\author{Manfred Bayer}
\affiliation{Experimentelle Physik 2, Technische Universität Dortmund,
  44221 Dortmund, Germany}


\maketitle

In this Supplemental Material, we provide a detailed description of
the methods used in the primary text.

\section{Periodic orbit theory}
\subsection{Semiclassical density of states}
In semiclassical theories~\cite{Berry76,Tomsovic95,Gut90} the density
of states is given by
\begin{equation}
  \varrho_{\mathrm{sc}}(E) = \varrho_0(E)
  + \sum_{\mathrm{po}} {\cal A}_{\mathrm{po}}
  \cos( S_{\mathrm{po}}(E)/\hbar- \frac{\pi}{2}\mu_{\mathrm{po}}) \, ,
\label{eq:rho_sc}
\end{equation}
with $\varrho_0(E)$ the average density of states.
The sinusoidal fluctuations of the density of states are related to
the periodic orbits of the classical system with ${\cal A}_{\mathrm{po}}$,
$S_{\mathrm{po}}$, and $\mu_{\mathrm{po}}$ the amplitude, action, and
Maslov index of the orbits, respectively.
In this treatment integrable and non-integrable systems must be
distinguished.

The dynamics of integrable systems is regular and the motion is
confined to tori characterized by action-angle variables $J_i$ and
$\vartheta_i$.
The action variables
\begin{equation}
    J_i=\frac{1}{2\pi} \oint_{{\cal C}_i} \boldsymbol{p} \mathrm{d}\boldsymbol{r}
\end{equation}
provide a set of constants of motion, where the integrals are evaluated
along the independent paths ${\cal C}_i$ on the torus obtained when
the angle variable $\vartheta_i$ is varied from $0$ to $2\pi$ while
all other angle variables $\vartheta_j$ with $j\ne i$ are kept constant.
The frequencies $\omega_i$ on the torus are given by
\begin{equation} 
  \omega_i= \frac{\partial H}{\partial J_i}. 
\end{equation}
When all frequency ratios are given by rational numbers the
trajectories are periodic on resonant tori, otherwise they are
quasi-periodic.
The periodic orbits can be characterized by integer winding numbers $M_i$ 
that are defined by the number of cycles along the paths ${\cal C}_i$ needed
until the periodic orbit returns to its initial position. 

For two-dimensional integrable systems the density of states is given
by the Berry-Tabor formula~\cite{Berry76,Tomsovic95}
\begin{align}
\begin{split}
  &\varrho_{\mathrm{sc}}(E) = \varrho_0(E)\\
  &+\frac{1}{\pi \hbar}\sum_{\boldsymbol{M}}
    \frac{\dd S_{\boldsymbol{M}} / \dd E}{\sqrt{\hbar M_2^3 \abs{g''_E}}}
    \cos(S_{\boldsymbol{M}}/\hbar -\frac{\pi}{2}\mu_{\boldsymbol{M}}
    -\frac{\pi}{4}) \, .
\end{split}
\label{eq:BT}
\end{align}
where $\boldsymbol{M}=(M_1,M_2)$ are the winding numbers of the
periodic orbits with classical action $S_{\boldsymbol{M}}$ and Maslov
index $\mu_{\boldsymbol{M}}$.
The amplitudes of the oscillating terms in Eq.~\eqref{eq:BT} involve
the periods $\dd S_{\boldsymbol{M}} / \dd E$ of the the periodic orbits,
the winding number $M_2$, and the second derivative of the function
$g_E$, which connects the two action variables via $J_2=g_E(J_1)$.

In non-integrable systems the periodic orbits are isolated.
The semiclassical density of states is then given by Gutzwiller's
trace formula~\cite{Gut90}
\begin{align}
\begin{split}
  &\varrho_{\mathrm{sc}}(E) = \varrho_0(E)\\
  &+\frac{1}{\pi \hbar}\sum_{\mathrm{po}}
    \frac{\dd S_{\mathrm{ppo}} / \dd E}{\sqrt{\abs{\det(\boldsymbol{M}_\mathrm{po}-\boldsymbol{1})}}}
    \cos(S_\mathrm{po}/\hbar-\frac{\pi}{2}\mu_{\mathrm{po}}) \, ,
\end{split}
    \label{eq:gutz}
\end{align}
where $S_{\mathrm{ppo}}$ is the action of the primitive periodic
orbit, $\boldsymbol{M}_{\mathrm{po}}$ is the monodromy matrix that
describes the stability properties of the periodic orbit in the
directions orthogonal to the orbit, and $\mu_{\mathrm{po}}$ is the
Maslov index.

\subsection{Scaling technique}
For certain systems it is possible to introduce a scaling of the
coordinates and momenta such that the phase space structure of the
classical dynamics no longer depends on the energy $E$ or an
appropriately chosen scaling parameter.
Examples are billiard systems~\cite{Cvi89} or the hydrogen atom in a
magnetic field~\cite{Hol88,Fri89,Main1999a}.
In such systems a scaling parameter
\begin{equation}
  w = \frac{1}{\hbar_{\mathrm{eff}}} = \frac{n_{\mathrm{eff}}}{\hbar}
\end{equation}
can be introduced such that the shape of the classical orbits does not
depend on $w$ and the classical action of periodic orbits depend
linearly on $w$, i.e.,
\begin{equation}
  S_{\mathrm{po}}(w)/\hbar = \tilde S_{\mathrm{po}} w \, ,
\label{eq:scaledAction}
\end{equation}
with $\tilde S_{\mathrm{po}}$ the scaled action of the periodic orbit.
The scaling parameter $w$ plays the role of an inverse effective
Planck constant $\hbar_{\mathrm{eff}}$ or an effective quantum number
$n_{\mathrm{eff}}$.

Changing the external parameter from energy to $w$ and inserting
Eq.~\eqref{eq:scaledAction} into the general form of the semiclassical
trace formula~\eqref{eq:rho_sc} the semiclassical density of states
can be written as a Fourier series 
\begin{equation}
  \varrho_{\mathrm{sc}}(w) = \varrho_0(w)
  +\Re \sum_{\mathrm{po}} {\cal A}_{\mathrm{po}}
    \exp(i\tilde S_{\mathrm{po}} w) \, ,
\label{eq:rho_series}
\end{equation}
with constant periodic orbit parameters, where we have included
the phase given by the Maslov index in the complex amplitude
${\cal A}_{\mathrm{po}}$.

\subsection{Scaled recurrence spectra}
In the semiclassical density of states~\eqref{eq:rho_series} of a
scaled system the complete set of periodic orbits lead to a
superposition of sinusoidal fluctuations.
The contribution of individual orbits to the spectrum is therefore not
visible directly, but can be extracted by a Fourier transform of the
spectrum from energy to time domain, or in the scaled picture from the
$w$ to the scaled action domain.
The scaled semiclassical recurrence spectrum
\begin{equation}
  C_{\mathrm{sc}}(\tilde S)
  = \sum_{\mathrm{po}} {\cal A}_{\mathrm{po}} \delta(\tilde S - \tilde S_\mathrm{po} ),
\label{eq:rec}
\end{equation}
then exhibits peaks at the positions of the scaled actions
$\tilde S_\mathrm{po}$ with amplitudes ${\cal A}_{\mathrm{po}}$.

The semiclassical recurrence spectrum~\eqref{eq:rec} can be directly
compared with its quantum mechanical counterpart.
For scaled systems the Schrödinger equation describes eigenstates at
discrete values $w=w_k$ of the scaling parameter.
The Fourier transform of the quantum mechanical density of states
\begin{equation}
  \varrho_{\mathrm{qm}}(w) = \sum_k \delta(w-w_k)
\label{eq:rho_qm}
\end{equation}
can be calculated analytically and yields the quantum mechanical
recurrence spectrum
\begin{align}
  C_{\mathrm{qm}}(\tilde S)
  &= \frac{1}{2\pi} \int \varrho_{\mathrm{qm}}(w) e^{-i\tilde S w} \dd w \nonumber \\
  &= \frac{1}{2\pi} \sum_k e^{-i\tilde S w_k} \, .
\label{eq:rec_qm}
\end{align}
By comparing the quantum mechanical result~\eqref{eq:rec_qm} with
Eq.~\eqref{eq:rec} it becomes clear that also the quantum mechanical
recurrence spectrum should provide recurrence peaks at positions
$\tilde S$ given by the scaled actions of the periodic orbits of the
corresponding classical system.
This allows for the interpretation of the quantum spectra in terms of
classical orbits.
More details on the application of the semiclassical approach and the
scaling technique to excitons in cuprous oxide are given below.

\section{Classical and semiclassical description of excitons in cuprous oxide}
\subsection{Scaling of the classical dynamics}
Excitons in semiconductors like Cu$_2$O are atom-like states formed by
an electron in the conduction band and a positively charged hole in
the valence band.
Neglecting all details of the band structure an exciton can be
described in a simple hydrogenlike model by the Hamiltonian
\begin{equation}
  H = E_{\mathrm{g}} + \frac{\gamma'_1}{2m_0} \boldsymbol{p}^2
  - \frac{e^2}{4\pi\varepsilon_0\varepsilon|\boldsymbol{r}|} \, .
\label{eq:H_hyd}
\end{equation}
The first term $E_{\mathrm{g}}=2.17208\,$eV is the gap energy
between the uppermost valence band and the lowest conduction band
\cite{GiantRydbergExcitons} and
$\boldsymbol{r}$ and $\boldsymbol{p}$ are the
relative coordinates and corresponding momenta.
Here, an appropriate scaling to make the dynamics energy-independent
is given by
\begin{equation}
  \boldsymbol{r} = n_{\mathrm{eff}}^2\tilde{\boldsymbol{r}} \, , \;
  \boldsymbol{p} = \frac{1}{n_{\mathrm{eff}}}\tilde{\boldsymbol{p}} \, ,
\label{eq:r_scal}
\end{equation}
where we introduced $n_{\mathrm{eff}}$ via
$n_{\mathrm{eff}}^2\equiv E_{\mathrm{Ryd}}/(E_{\mathrm{g}}-E)$
with $E_{\mathrm{Ryd}}=13.6$eV$/(\gamma_1' \varepsilon^2)\approx 87$meV the Rydberg energy of cuprous oxide.
After subtracting $E_\mathrm{g}$ and multiplying with $n_{\mathrm{eff}}^2$
the scaled Hamiltonian reads
\begin{equation}
\tilde H= n_{\mathrm{eff}}^2(H-E_{\mathrm{g}})=\frac{\gamma'_1}{2m_0} \tilde{\boldsymbol{p}}^2
  - \frac{e^2}{4\pi\varepsilon_0\varepsilon|\tilde{\boldsymbol{r}}|}=-E_{\mathrm{Ryd}} \, 
\end{equation}
and therefore the classical dynamics becomes independent of $n_{\mathrm{eff}}$.
The connection between non-scaled action $S$ and scaled action $\tilde
S$ is then given by
\begin{equation}
  S_{\mathrm{po}}(n_{\mathrm{eff}}) = \tilde{S}_{\mathrm{po}} n_{\mathrm{eff}} \, .
\label{scaled_action_excitons}
\end{equation}
The semiclassical density of states therefore takes the
form~\eqref{eq:rho_series} when setting $w=n_{\mathrm{eff}}/\hbar$,
and the recurrence spectrum exhibits $\delta$ peaks at scaled actions
$\tilde S_{\mathrm{po}}$ of multiple repetitions of Keplerian orbits
corresponding to the hydrogenlike Coulomb problem.

The true exciton dynamics is much more complicated.
A full description of excitons in cuprous oxide needs to include the
cubic band structure of the crystal.
This requires the replacement of the kinetic term in
Eq.~\eqref{eq:H_hyd} by
\begin{equation}
  \frac{\gamma'_1}{2m_0} \boldsymbol{p}^2 \to
  H_{\mathrm{kin}} (\boldsymbol{p},\boldsymbol{\hat{I}},\boldsymbol{\hat{S}}_{\mathrm{h}})
  + H_\mathrm{SO}(\boldsymbol{\hat{I}},\boldsymbol{\hat{S}}_{\mathrm{h}}),
\label{eq:kinetic}
\end{equation}
with the vector operators $\boldsymbol{\hat{I}}$, $\boldsymbol{\hat{S}}_{\mathrm{h}}$
for angular momenta $I=1$ and $S_\mathrm{h}=1/2$.
The kinetic terms can be split into the hydrogenlike term plus
corrections due to the band structure
$H_{\mathrm{band}}(\boldsymbol{p},\boldsymbol{\hat{I}},\boldsymbol{\hat{S}}_{\mathrm{h}})$, i.e.
\begin{align}
  &H_{\mathrm{kin}} (\boldsymbol{p},\boldsymbol{\hat{I}},\boldsymbol{\hat{S}}_{\mathrm{h}})
    + H_\mathrm{SO}(\boldsymbol{\hat{I}},\boldsymbol{\hat{S}}_{\mathrm{h}})  = \frac{\gamma'_1}{2m_0} \boldsymbol{p}^2+H_\mathrm{band}(\boldsymbol{p},\boldsymbol{\hat{I}},\boldsymbol{\hat{S}}_{\mathrm{h}})\nonumber\\[1ex]
  & = \frac{\gamma'_1}{2m_0} \boldsymbol{p}^2
    +\frac{1}{2\hbar^2m_0} \big[4\gamma_2\hbar^2\boldsymbol{p}^2
  -6\gamma_2(p^2_1\boldsymbol{\hat{I}}^2_1+{\rm c.p.})\nonumber\\[1ex]
  &  -12\gamma_3(\{p_1,p_2\}\{\hat{I}_1,\hat{I}_2\}+{\rm c.p.})
  -12\eta_2(p^2_1\hat{I}_1\hat{S}_{\rm h1}+{\rm c.p.})\nonumber\\[1ex]
  &  +2(\eta_1+2\eta_2)\boldsymbol{p}^2(\boldsymbol{\hat{I}}\cdot\boldsymbol{\hat{S}}_{\rm h})-12\eta_3(\{p_1,p_2\}(\hat{I}_1\hat{S}_{\rm h2}
  \nonumber\\[1ex]
  &+\hat{I}_2\hat{S}_{\rm h1})+{\rm c.p.})\big] 
    +H_\mathrm{SO}(\boldsymbol{\hat{I}},\boldsymbol{\hat{S}}_{\mathrm{h}}) \, .
\label{eq:H_kin}
\end{align}
We here neglect the central-cell corrections~\cite{frankevenexcitonseries,
Farenbruch2020a,Rommel2021a},
since these corrections including the electron-hole exchange 
interaction become import at very small electron-hole distance $r \to 0$.
Therefore, mostly even exciton states with a high percentage of an S
state are affected.
The number of S states compared to the total number of exciton states
deceases strongly with increasing principal quantum number $n$.
We therefore assume that central-cell corrections can be neglected in
the semiclassical limit.
This statement is supported by our classical computations, where we
observe that the classical exciton orbits very rarely reach the region
with small electron-hole distance $r \to 0$.

The band structure terms included in 
$H_{\mathrm{kin}}(\boldsymbol{p},\boldsymbol{\hat{I}},\boldsymbol{\hat{S}}_{\mathrm{h}})$
depend quadratically on components $p_i$ of the momentum, and
therefore the scaling transformation~\eqref{eq:r_scal} can still be
applied to obtain an energy-independent classical dynamics.
However, this does not hold for the spin-orbit coupling term
\begin{equation}
  H_{\mathrm{SO}}=\frac{2}{3}\Delta
  \left(1+\frac{1}{\hbar^2}\boldsymbol{\hat{I}}\cdot\boldsymbol{\hat{S}}_{\mathrm{h}}\right)\, .
\label{eq:H_SO}
\end{equation}
which does not depend on the momentum $\boldsymbol{p}$ and thus
depends on $n_{\mathrm{eff}}$ after multiplication of the total
Hamiltonian by $n_{\mathrm{eff}}^2$.
To recover the scaling property of the system we introduce a scaled
version of the spin-orbit coupling by the replacement
\begin{equation}
  \Delta \to \tilde \Delta = \frac{n_0^2}{n_{\mathrm{eff}}^2} \Delta \, .
\label{eq:Delta_scal}
\end{equation}
The dependence of the total Hamiltonian on $n_{\mathrm{eff}}$ can now
be completely removed by the scaling, but the scaled classical
dynamics depends on the constant factor $n_0^2$, which determines the
strength of the spin-orbit coupling.
The classical dynamics of the scaled system corresponds exactly to
that of the non-scaled crystal at energy
$E=E_{\mathrm{g}}-E_{\mathrm{Ryd}}/n_{\mathrm{eff}}^2$ with $n_{\mathrm{eff}}=n_0$.

\subsection{Adiabatic approach for the exciton dynamics}
To calculate the classical dynamics of the yellow exciton series we
resort to the adiabatic approach introduced in Ref.~\cite{Ertl2020}.
Thereby we assume that the fast motion of the spin degrees of freedom
reacts instantaneously to a new configuration in the relative
coordinates.
The timescale of the spin dynamics is determined by the spin-orbit
coupling $\Delta$ which is large already compared to the spacing of
adjacent Rydberg states $2E_{\mathrm{Ryd}}/n^3$ for principle quantum
numbers $n\geq3$.
Diagonalizing the band structure part $H_\mathrm{band}(\boldsymbol{p},\boldsymbol{\hat{I}},\boldsymbol{\hat{S}}_{\mathrm{h}})$
in a basis for quasispin and hole spin $\ket{m_I, m_{S_{\rm h}}}$ with $m_I$ and $m_{S_{\mathrm{h}}}$ the
magnetic quantum numbers of the quasispin and hole spin, leads to three twofold degenerate
energy surfaces in momentum space $W_i(\boldsymbol{p},n_0)$.
The lowest of those energy surfaces can be assigned  to the yellow
exciton series.
To interpret the fine-structure splitting of the yellow exciton series in cuprous oxide
in terms of the classical exciton orbits, we have to choose the corresponding energy surface $W_1(\boldsymbol{p},n_0)$.
The Hamilton function is then given by
\begin{equation}
  H = \frac{\gamma'_1}{2m_0} \boldsymbol{p}^2 + W_1(\boldsymbol{p},n_0)
  - \frac{e^2}{4\pi\varepsilon_0\varepsilon|\boldsymbol{r}|} \, .
\label{eq:H_surf}
\end{equation}
The classical exciton orbits are calculated by numerically integrating
Hamilton's equations of motion
\begin{equation}
  \dot r_i = \frac{\gamma'_1}{m_0} p_i + \frac{\partial
              W_1(\boldsymbol{p},n_0)}{\partial p_i} \; , \quad
  \dot p_i = -\frac{e^2}{4\pi\varepsilon_0\varepsilon}
              \frac{r_i}{|\boldsymbol{r}|^3} \, .
\end{equation}
Periodic orbits can be found by varying the initial conditions.
We find  periodic orbits confined to one- to three-dimensional tori.
The one- and two-dimensional orbits are located in the two distinct 
symmetry planes of the crystal with $O_\mathrm{h}$ symmetry. 
The phase space exhibits mostly regular tori surrounding an elliptical
fixed point given by a nearly circular orbit.
The two-dimensional orbits show a secular motion of Kepler ellipses
which can be characterized by two winding numbers $M_1$ and $M_2$,
whereby $M_1$ gives the number of Kepler ellipses and $M_2$ describes
the secular motion by giving the number of circulations on the torus.
For the three-dimensional orbits a third winding number needs to be included.
$M_2$ then describes the secular motion in $\varphi$ direction and
$M_3$ describes the secular motion in $\vartheta$ direction.

Additional periodic orbit parameters,
which are needed for the calculations of the semiclassical amplitudes,
can be calculated by integrating the corresponding equations of motion
simultaneously.
The equation of motion for the action is given by
\begin{equation}
  \frac{\mathrm{d}S}{\mathrm{d}t}
  = \boldsymbol{p} \frac{\mathrm{d}\boldsymbol{r}}{\mathrm{d}t} \, .
\end{equation}
The stability properties of a periodic orbits are determined by the
symplectic stability matrix $\boldsymbol{M}_{\mathrm{po}}$, which is
obtained by integrating
\begin{equation}
\frac{\mathrm{d}}{\mathrm{d}t} \boldsymbol{M} = \boldsymbol{J} \frac{\partial^2 H}{\partial \boldsymbol{\gamma} \partial \boldsymbol{\gamma}} \boldsymbol{M}\, , \, \mathrm{with} \, 
\boldsymbol{J} =\begin{pmatrix} \phantom{-}\boldsymbol{0} & \boldsymbol{1}\\ -\boldsymbol{1} & \boldsymbol{0} \end{pmatrix}\, , \,
\boldsymbol{\gamma}=\begin{pmatrix} \boldsymbol{r} \\ \boldsymbol{p} \end{pmatrix}\, ,
\end{equation}
over one period of the orbit with the initial condition
$\boldsymbol{M}(0)=\boldsymbol{1}$.
Due to the symplectic structure of the stability matrix its eigenvalues
appear in pairs $\lambda_i$ and $1/ \lambda_i$.

\begin{figure}
\includegraphics[width=\columnwidth]{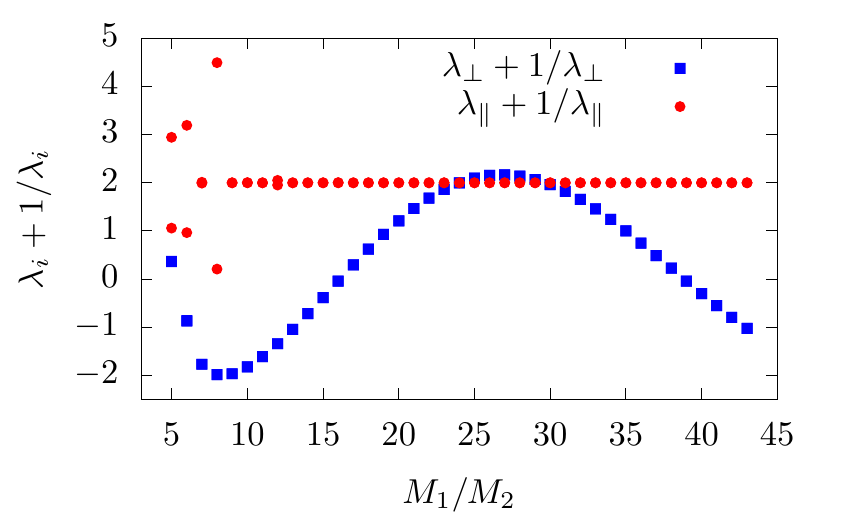}
\caption{Sum $\lambda_i+1/\lambda_i$ for the two-dimensional
  orbits at $n_0=5$ in the   plane normal to $[100]$. For the majority
  of orbits the deviation from   stability eigenvalues $\lambda_i=1$
  cannot be resolved.}
\label{fig:Stability_plane}
\end{figure}

For the nearly circular orbits all stability eigenvalues differ
significantly from one.
The direction out of the plane is stable for the orbit in the plane
normal to $[100]$ and unstable for the orbit in the plane normal to
$[1\bar{1}0]$.
This behavior also holds for most of the two-dimensional orbits in the
different symmetry planes.
The direction out of the plane normal to $[1\bar{1}0]$ is unstable.
By contrast, the stability out of the plane normal to $[100]$ shows
stable behavior for most of the orbits in the symmetry plane.
In the planes two partner orbits, one stable and one unstable orbit
can be found in accordance with the Poincaré-Birkhoff theorem.
However, the deviation of the stability eigenvalues from
one can only be resolved numerically for the orbits with the lowest
ratios $M_1/M_2$.
This can be seen in Fig.~\ref{fig:Stability_plane} where the sums
$\lambda_i+1/\lambda_i$ are plotted over the ratio $M_1/M_2$ for the
planar orbits normal to $[100]$ for a fixed value $n_0=5$.
$\lambda_\parallel$ describes the stability of the orbit in the plane and
$\lambda_\perp$ the stability out of the plane. 
The majority of two-dimensional orbits exhibit one pair of stability eigenvalues with
$\lambda_\parallel=1$. 
The same behavior can also be observed for the two-dimensional orbits
in the other symmetry plane and for the three-dimensional orbits.

\subsection{Calculation of the periodic orbit amplitudes}
In this work we present results for $n_0=5$.
This choice fixes the classical dynamics corresponding to
the scaled exciton spectrum to that belonging to the exciton state at
principal quantum number $n=5$ in the non-scaled crystal.
The value $n_0=5$ has been selected for two reasons.
On the one hand, the secular motion of the classical exciton orbits
in similar phase space regions
becomes slower with increasing $n_0$ and non-trivial structures in the
recurrence spectra occur only at higher values of the scaled action~\cite{SchumacherMsc}.
On the other hand,
lower values $n_0<3$ belong to regions where the adiabatic approach
introduced in Ref.~\cite{Ertl2020} is no longer valid.

The one-dimensional orbits are isolated and their stability
eigenvalues therefore differ from one.
Their contribution to the density of states can be calculated by applying 
Gutzwiller's trace formula \eqref{eq:gutz}. 
However stability eigenvalues equal to one are typical for integrable systems
where the density of states can be described by the Berry-Tabor formula.
To calculate semiclassical amplitudes for two- and three-dimensional orbits
we therefore replace the contributions of the directions belonging to stability 
eigenvalues $\lambda_i=1$ in Gutzwiller's trace formula by the
Berry-Tabor formula~\eqref{eq:BT} for a two-dimensional integrable system.
The contribution of the third direction is then given by a factor
$1/\sqrt{\abs{\lambda_{\mathrm{po}}+1/ \lambda_{\mathrm{po}}-2}}$.
The amplitudes for the two- and three-dimensional orbits then read
\begin{equation}
|{\cal A}_{\mathrm{po}}|=
\frac{1}{\pi \hbar}
      \frac{1}{\sqrt{\abs{\lambda_{\mathrm{po}}+1/ \lambda_{\mathrm{po}}-2}}}
      \frac{\tilde{S}_{\mathrm{po}}}{\sqrt{\hbar M_2^3 \abs{g''_E}}} \, .
\label{eq:A_2D_r}
\end{equation}
To calculate these amplitudes the action variables $J_1$ and $J_2$ are needed.
In a two-dimensional integrable system the action of a periodic orbit is given by
\begin{equation}
  S_{\boldsymbol{M}}=2\pi (M_1 J_1 + M_2 J_2) \, .
\label{eq:Action_int}
\end{equation}
The action variables can then be calculated by differentiating
Eq.~\eqref{eq:Action_int} by the corresponding winding number $M_i$.
\begin{figure}
\includegraphics[width=\columnwidth]{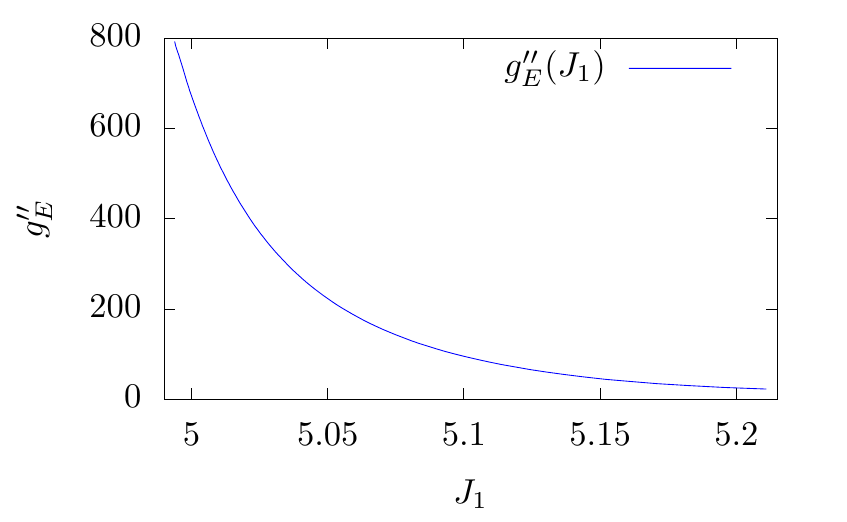}
    \caption{Second derivative of the function $J_2=g_E(J_1)$ with respect to $J_1$ at $n_0=5$.
    The orbits in the plane normal to $[100]$ produce a steady curve which can be 
    interpolated for orbits with intermediate values for $J_1$ and $J_2$.
    All quantities are given in exciton-Hartree units which are obtained by setting $\hbar = e = m_{\mathrm{0}} / \gamma_1'
= 1 / (4\pi\varepsilon_{\mathrm{0}} \varepsilon) = 1$.        }
    \label{fig:gss}
\end{figure}
For two-dimensional motion $M_1$ describes the number of Kepler
ellipses, whereas $M_2$ describes the secular motion of an orbit.
To describe the three-dimensional orbits within the same approach we
treat them in an effective two-dimensional model by combining the
contribution of the secular motion in $\vartheta$ direction described
by $M_2$ and the secular motion in $\varphi$ direction described by
$M_3$ to give an overall contribution of the secular motion, which is
then described by $\tilde{M}_2=GCD(M_2,M_3)$. 
With the action variables $J_1$ and $J_2$ the function $J_2=g_E(J_1)$
can be calculated.
In Fig.~\ref{fig:gss} the second derivative  $g_E''$  with respect to
$J_1$ is shown for the two-dimensional orbits in the plane normal to
$[100]$ for $n_0=5$.
The amplitudes in Eq.~\eqref{eq:A_2D_r} are calculated by choosing  
the $g_E''$ of the corresponding orbit.

\section{Scaled quantum mechanical exciton spectra}
In the quantum mechanical case the canonical commutation relations
\begin{equation}
  [ \hat{r}_i, \hat{p}_j ] = i \hbar \delta_{ij}
\label{eq:com}
\end{equation}
must be satisfied.
Inserting the scaled variables given in Eq.~\eqref{eq:r_scal} yields
\begin{equation}
  [ \hat{\tilde r}_i, \hat{\tilde p}_j ] = i \frac{\hbar}{n_{\mathrm{eff}}} \delta_{ij} \, ,
\label{eq:com_scal}
\end{equation}
where the Planck constant is replaced by an effective Planck constant
$\hbar_{\mathrm{eff}} = \hbar /n_{\mathrm{eff}}$.
Using the corresponding operators in coordinate space given by
\begin{equation}
\hat{\tilde{\boldsymbol{r}}} = \tilde{\boldsymbol{r}}\, , \;
\hat{\tilde{\boldsymbol{p}}} = -i\hbar_{\mathrm{eff}}\nabla_{{\tilde r}}\, ,
\label{eq:r_scal_operator}
\end{equation}
the Schrödinger equation for cuprous oxide can be transformed to the
generalized eigenvalue problem 
\begin{align}
  &\left[\frac{e^2}{4\pi\varepsilon_0\varepsilon|\boldsymbol{r}|} -n_0^{2}H_\text{SO}(\boldsymbol{\hat{I}},\boldsymbol{\hat{S}}_{\mathrm{h}}) - E_\mathrm{Ryd} \right] |\Psi \rangle \nonumber \\ 
  &= \frac{\hbar^2}{n_{\mathrm{eff}}^2} H_{\mathrm{kin}}
  (-i \nabla_{\tilde r},\boldsymbol{\hat{I}},\boldsymbol{\hat{S}}_{\mathrm{h}}) 
  |\Psi \rangle
\label{eq:H_scal}
\end{align}
for the effective quantum number $n_{\mathrm{eff}}$.

\subsection{Numerical solution of the generalized eigenvalue problem}
To solve the generalized eigenvalue problem~\eqref{eq:H_scal} we use
the basis set
\begin{equation}
  \ket{\Pi} = \ket{N,L,J,F,M_F}\, .
  \label{eq:basis_set}
\end{equation}
Here, $N$ is the radial quantum number which is connected to the
principal quantum number $n$ via $N=n-L-1$ and $L$ the angular
momentum quantum number.
The quasispin $I$ and the hole spin $S_{\mathrm{h}}$ are coupled via
the spin-orbit coupling term $H_\mathrm{SO}$ to the effective hole
spin $J$.
The total angular momentum quantum number is denoted by $F = J + L$
and its $z$ component by $M_F$.
As we neglect the central-cell corrections affecting the even-parity
states, the electron spin $S_e$ and its $z$ component $M_{S_e}$ are
good quantum numbers.
They only result in additional degeneracies of the eigenstates and
thus are not included in the basis~\eqref{eq:basis_set}.

For the radial part of the basis states we use the Coulomb-Sturmian
functions \cite{frankevenexcitonseries}
\begin{equation}
  U_{NL}(\tilde{r}) = N_{NL}~(2\rho)^L e^{-\rho}L_N^{2L+1}(2\rho)\, ,
  \label{eq:radial_part_CS}
\end{equation}
with the associated Laguerre polynomials $L_N^{2L+1}(x)$,
$\rho=\tilde{r}/\alpha$, and $\alpha$ an arbitrary positive
real-valued scaling parameter.
The Coulomb-Sturmian functions form a discrete and complete set in
$L^2(\mathbb{R}^3)$ independent of the scaling parameter $\alpha > 0$
\cite{CoulombSturmForNuclear, klahn1977convergence_I, klahn1977convergence_II}.
The normalization factor $N_{NL}$ reads
\begin{equation}
  N_{NL} = \frac{2}{\sqrt{\alpha^3}}\left[ \frac{N!}{(N+2L+1)! (N+L+1)!} \right]^{\frac{1}{2}}\, .
  \label{eq:CS_normalization}
\end{equation}
Note that the functions~\eqref{eq:radial_part_CS} are not orthogonal.
For the angular parts of the basis states we use the spherical harmonics.
For the eigenfunctions we use the ansatz
\begin{equation}
	\ket{\Psi} = \sum_{N,L,J,F,M_F} c_{N,L,J,F,M_F} \ket{\Pi}\, ,
	\label{eq:ansatz}
\end{equation}
with coefficients $c$. 
Using the ansatz \eqref{eq:ansatz} we can write the generalized
eigenvalue problem in matrix form
\begin{equation}
  \boldsymbol{A}\boldsymbol{c} = \lambda \boldsymbol{B}\boldsymbol{c}\, ,
  \label{eq:generalized_eigenvalue_problem}
\end{equation}
with symmetric matrices $\boldsymbol{A}$ and $\boldsymbol{B}$, the
solution vector $\boldsymbol{c}$, and the eigenvalues
$\lambda = 1/n_{\mathrm{eff}}^2$.
The matrix elements of $\boldsymbol{A}$ and $\boldsymbol{B}$,
\begin{align}
  A_{\tilde{\Pi}\Pi} &= \bra*{\tilde{\Pi}}\left[\frac{e^2}{4\pi\varepsilon_0\varepsilon|\boldsymbol{r}|} -n_0^{2}H_\text{SO}(\boldsymbol{\hat{I}},\boldsymbol{\hat{S}}_{\mathrm{h}}) - E_\mathrm{Ryd} \right]\ket{\Pi}\, ,\nonumber\\
  B_{\tilde{\Pi}\Pi} &= \bra*{\tilde{\Pi}}H_{\mathrm{kin}}
  (-i \nabla_{\tilde r},\boldsymbol{\hat{I}},\boldsymbol{\hat{S}}_{\mathrm{h}})\ket{\Pi}
  \label{eq:definition_matrices_scaled}
\end{align}
can be obtained analytically similar as in Ref.~\cite{ImpactValence}.
They have been adjusted to the basis set \eqref{eq:basis_set} using
Clebsch–Gordan coefficients.
The matrix equation~\eqref{eq:generalized_eigenvalue_problem} 
is set up using the basis set~\eqref{eq:basis_set} with quantum
numbers
\begin{align}
	L &= 0, \dots, n-1\nonumber\, ,\\
	J &= 1/2, 3/2\nonumber\, ,\\
	F &= |L-J|, \dots, L+J\nonumber\, ,\\
	M_F &= -F, \dots, +F\, .
	\label{eq:quantum_numbers_possible_values}
\end{align}

As parity is a good quantum number, the generalized eigenvalue problem
decomposes into two blocks with even and odd parity which corresponds
to even and odd angular momentum $L$. Those blocks further decompose
into four blocks because of the cubic symmetry of the kinetic part of
the Hamiltonian $H_\mathrm{kin}$ and the quantization axis chosen
parallel to one of the principal axis of the crystal. More precisely,
as a rotation of $90^\circ$ about a four-fold symmetry axis of the
group $O_\mathrm{h}$ does not change the Hamiltonian and one of those
axis coincides with the quantization axis it follows
\begin{align}
	\bra{\Pi'}e^{\frac{iF_z \pi/2}{\hbar}} H_\mathrm{kin} e^{\frac{-iF_z \pi/2}{\hbar}}\ket{\Pi} &= e^{\frac{i(M_F'-M_F) \pi/2}{\hbar}} \bra{\Pi'} H_\mathrm{kin} \ket{\Pi}\nonumber\\
	&= \bra{\Pi'} H_\mathrm{kin} \ket{\Pi}\, ,
	\label{eq:rotation_about_4fold_axis}
\end{align}
which is only fulfilled if $M_F'-M_F = 4m$ with $m \in \mathbb{Z}$. Thus only states with $M_F = M_F' \mod 4$ couple.
The four blocks correspond to the four equivalence classes of $M_F \mod 4$
\begin{align}
	\left[ 1/2 \right] &= \left\{ \dots, -15/2, -7/2, 1/2, 9/2, 17/2, \dots \right\}\, ,\nonumber\\
	\left[ 3/2 \right] &= \left\{ \dots, -13/2, -5/2, 3/2, 11/2, 19/2, \dots \right\}\, ,\nonumber\\
	\left[ 5/2 \right] &= \left\{ \dots, -11/2, -3/2, 5/2, 13/2, 21/2, \dots \right\}\, ,\nonumber\\
	\left[ 7/2 \right] &= \left\{ \dots, -9/2, -1/2, 7/2, 15/2, 23/2, \dots \right\}\, .
	\label{eq:blocks}
\end{align}
The symmetry of a state can be specified by the irreducible representations $\Gamma_i$ of the cubic $O_\mathrm{h}$ group which determine the transformation behavior under symmetry operations of the group.
Because the angular momentum $F$ is half-integer only the irreducible representations $\Gamma_6 ,\Gamma_7$ and $\Gamma_8$ appear \cite{koster1963properties}.
The $\Gamma_8$ states are four-fold degenerate and one of those degenerate states appear in each block of Eq.~\eqref{eq:blocks}. In the blocks $[1/2]$ and $[7/2]$ additionally only $\Gamma_7$ states, and in the blocks $[3/2]$ and $[5/2]$ additionally only $\Gamma_6$ states appear. Those $\Gamma_6$ and $\Gamma_7$ states are globally two-fold degenerate with one of the degenerate states in each block they occur in. 
Thus the block structure together with the degeneracy of the states can be used to assign the irreducible representation to a given state.

With the symmetry plane orthogonal to the four-fold axis along the
$z$ direction it follows that 
\begin{align}
	&\bra{N',L',J',F',M_F'} H_\mathrm{kin} \ket{N,L,J,F,M_F}\nonumber\\
	&= \bra{N',L',J',F',-M_F'} H_\mathrm{kin} \ket{N,L,J,F,-M_F}\, ,
	\label{eq:z-refection-symmetry}
\end{align}
and thus for the matrices $\boldsymbol{B}$ defined in Eq.~\eqref{eq:definition_matrices_scaled}
\begin{align}
	\boldsymbol{B}_{M_F \in \left[ 1/2 \right]} &\sim \boldsymbol{B}_{M_F \in \left[ 7/2 \right]}\, ,\nonumber\\
	\boldsymbol{B}_{M_F \in \left[ 3/2 \right]} &\sim \boldsymbol{B}_{M_F \in \left[ 5/2 \right]}\, ,
	\label{eq:equal_blocks}
\end{align}
where $\sim$ denotes similar matrices and thus their spectra are equal. They are similar because they can be transformed into each other by a basis transformation as only the arrangement of the basis vectors has to be changed.
This explains the two-fold degeneracy of the $\Gamma_6$ and $\Gamma_7$ states.
Thus the generalized eigenvalue problem \eqref{eq:generalized_eigenvalue_problem} decomposes into four independent blocks, two for even and odd parity, respectively.
This block structure is exploited to accelerate the numerical diagonalization using the LAPACK routine DSYGVX \cite{lapackuserguide3}. For the latter we need a finite basis and therefore set upper bounds $n_\mathrm{max}$ and $L_\mathrm{max}$ to the principal quantum number $n$ and the angular momentum quantum number $L$, respectively. They are chosen sufficiently large to achieve convergence. The convergence parameter $\alpha$ is the same for all basis functions and can be used to enhance convergence. Our truncated basis set is therefore uniquely defined by the triple $(\alpha, n_\mathrm{max}, L_\mathrm{max})$.

\subsection{Calculation of quantum mechanical recurrence spectra}
The Fourier transform of the scaled quantum density of states 
\begin{equation}
	\varrho(n_{\mathrm{eff}}) = \sum_{k}\delta(n_{\mathrm{eff}} - n_{\mathrm{eff},k})\, ,
	\label{eq:scaled_qm_DOS}
\end{equation}
where $n_{\mathrm{eff},k}$ denotes the $k$th eigenvalue of
Eq.~\eqref{eq:generalized_eigenvalue_problem}, should result in
$\delta$ peaks at the scaled actions $\tilde S_{\mathrm{po}}$ and their repetitions.
This is only valid for an infinite spectrum. 
For the finite one, resulting from the diagonalization of the
truncated generalized eigenvalue problem, the $\delta$ peaks are
broadened and side peaks occur.
This can be seen as follows.
The Fourier transform of the finite spectrum corresponds to that of
the infinite one multiplied by a rectangular window function.
With the convolution theorem this is equivalent to a convolution of
the Fourier transformed infinite spectrum with 
\begin{equation}
	\frac{\sin(\Delta n_\mathrm{eff} \tilde S_{\mathrm{po}}/(2\hbar))}{\pi \tilde S_{\mathrm{po}}/\hbar} e^{-i n_\mathrm{eff}^0 \tilde S_{\mathrm{po}}/\hbar}\, ,
\end{equation}
where $\Delta n_\mathrm{eff}$ is the size of the window, i.e., the
length of the finite spectrum, and $n_\mathrm{eff}^0$ the center of
the spectrum.
To suppress the artificial side peaks a Gaussian window function
\begin{equation}
  w(n_\mathrm{eff}) \equiv \exp(-\frac{(n_\mathrm{eff}-n_\mathrm{eff}^0)^2}{2\sigma^2})
\end{equation}
is introduced with $\sigma = \Delta n_\mathrm{eff}/6$. Unfortunately
this further broadens the main peaks.
For our calculations we used $n_\mathrm{eff}^0=15$ and $\Delta n_\mathrm{eff}=30$.
As the quantum spectrum is a sum of delta distributions the Fourier
transform can be carried out analytically
\begin{align}
	&\hat{C}(\tilde S) = \frac{1}{2\pi} \sum_{k=1}^{k_\mathrm{max}} \int_{\mathbb{R}} w(n_\mathrm{eff}) \delta(n_\mathrm{eff} - n_{\mathrm{eff},k}) e^{-i n_\mathrm{eff} \tilde S/\hbar}~\mathrm{d}n_\mathrm{eff}\nonumber\\
	&= \frac{1}{2\pi} \sum_{k=1}^{k_\mathrm{max}} w(n_{\mathrm{eff},k})\left[ \cos(n_{\mathrm{eff},k} \tilde S/\hbar) - i\sin(n_{\mathrm{eff},k} \tilde S/\hbar) \right]\, ,
\end{align}
where $k=1,\dots, k_{\max}$ numerates the eigenvalues
$n_{\mathrm{eff},k}$ of the finite spectrum.

\bibliography{paper}